\documentclass[12pt]{article}
\usepackage{graphics}
\begin{document}
\markright{Extremal black holes...}
\title{Extremal black holes and the limits of the third law}
  \author{Stefano Liberati$^*$, Tony
  Rothman$^\dagger$ and Sebastiano Sonego$^\ddagger$\\[2mm] {\small
    \it \thanks{liberati@sissa.it}~ International School for Advanced
    Studies, via Beirut 2-4, Trieste 34014, Italy;}\\ {\small \it INFN
    sezione di Trieste.}\\ {\small \it
    \thanks{trothman@titan.iwu.edu}~Dept.\ of Physics, Illinois
    Wesleyan University, Bloomington, IL 61702, USA.}\\ {\small \it
    \thanks{sebastiano.sonego@uniud.it}~Universit\`a di Udine, Via
    delle Scienze 208, 33100 Udine, Italy.}\\ } 
    \date{{\small 10 August 2000; \LaTeX-ed \today}\\
{\small \it Paper awarded of an ``honorable mention'' in
    the Annual Competition\\ of the Gravity Research Foundation for the
    year 2000.}}
\maketitle

\begin{abstract}

  Recent results of quantum field theory on a curved spacetime suggest
  that extremal black holes are not thermal objects and that the
  notion of zero temperature is ill-defined for them. If this is
  correct, one may have to go to a full semiclassical theory of
  gravity, including backreaction, in order to make sense of the third
  law of black hole thermodynamics. Alternatively it is possible that we
  shall have to drastically revise the status of extremality in black hole
  thermodynamics.

\end{abstract}

Since the early 1970s, relativists have recognized that a
remarkable analogy exists between the three laws of
thermodynamics and the behavior of black holes, behavior
that is encoded in the three laws of black hole dynamics.
Indeed, since Hawking's celebrated discovery of black hole
radiation \cite{Hawking75}, the general consensus has been
that the laws of black hole dynamics are no mere analogy, as
remarkable as the analogy may be, but precisely the laws of
thermodynamics applied to black holes. Historically, most
research has focused on the first and second laws, which
explicitly identify the surface area of a black hole with
its entropy. In the past few years, however, increasing
attention has been paid to the third law, and it may be the
third law that is pointing toward the limits of black hole
thermodynamics as currently understood.

The third law of ordinary thermodynamics admits two
different formulations, both due to Nernst. The ``entropic''
formulation states that the entropy of a system approaches a
constant, independent of all the macroscopic parameters, as
the temperature approaches absolute zero. According to the
Planck postulate, this constant can be set equal to zero.
The second formulation, the so-called ``unattainability''
law, states that it is impossible to reach absolute zero in
a finite number of steps. For common physical systems the
two versions are equivalent, although one can imagine
situations in which the equivalence breaks down~\cite{Land}.

For a black hole with mass $M$, charge $Q$ and specific
angular momentum $a$, the temperature is proportional to the
surface gravity $\kappa \propto (M^2 - Q^2 - a^2)^{1/2}$,
and the entropy, according to the Bekenstein-Hawking
formula, is simply one-fourth the area of the event horizon.
Clearly, the temperature vanishes when $M^2=Q^2+a^2$, the
``extremal'' limit. However, the area still depends on the
parameters $Q$ and $a$ as $\kappa \to 0$, which represents
an obvious violation of the entropic formulation of the
third law. On the other hand, it is now well known that in
black hole dynamics one cannot reach the extremal state by
any continuous, finite-time process~\cite{Israel}; therefore
the unattainability formulation of the third law appears
favored.

Unfortunately, direct evaluations of the entropy do not all
lead to the same results: The current status can be broadly
summarized as follows: Semiclassical calculations, which
take into account the peculiar global structure of the
extremal spacetime manifold, lead to a zero value for the
entropy of extremal black
holes~\cite{GK95,HHR95,HH96,LP97,KL99}. On the other hand,
string theory calculations, in which the entropy can be
explained on purely statistical grounds, find that extremal
solutions obey the Bekenstein-Hawking formula
(see~\cite{Horow97} for a comprehensive review). Both sets
of results, however, imply a violation of the entropic
formulation of the third law. Indeed, even if extremal black
holes had a vanishing entropy, zero is not the value to
which the entropy of nearly extremal black holes tends in
the limit $\kappa\to 0$. This lack of a good limit hints at
the existence of a discontinuity in the thermodynamical
behaviour of the two classes of objects.

Such considerations suggest that one should focus on the thermodynamic
nature of the extremal black holes, rather than try to fix the entropic
formulation of the third law~\cite{Wald97}. Along these lines, we have
studied the thermal nature of an ``incipient'' extremal black hole --- a
collaping spherical body with an exterior extremal Reissner-Nordstr\"om
(RN) metric~\cite{LRS00}.

The main result of our investigation is that incipient extremal
black holes do not behave as thermal objects at any time during
their history. In other words, they do not radiate as black
bodies and their emission spectrum, as well as the
expectation value and variance of their stress-energy-momentum
tensor, do not correspond, at any finite time, to the limit for
$\kappa\to 0$ of the non-extremal incipient black holes.

Then one can conclude that neither temperature nor thermodynamic
entropy can be associated with them. This conclusion evidently
would lead to an explanation of the third law of black hole
thermodynamics in a radical way: {\em The temperature of a black
hole cannot be reduced to zero because no black hole zero
temperature state exists\/}.

The steps toward demonstrating that such incipient extremal RN black holes
have no thermal properties are reasonably straightforward and can be
summarized here. (For a few details see the appendix and for a full
derivation see \cite{LRS00}.) One employs the so-called ``moving mirror''
analogy to model the collapsing body by a one-dimensional mirror moving in
two-dimensional Minkowski space \cite{FD77,Birrell82}. On the background
we assume a propagating quantum scalar field. Incoming scalar rays,
passing through the center of the star are reflected off the mirror to
become outgoing rays. Due to the motion of the mirror, one expects the In
and Out vacuum states to differ, leading to particle production at late
times (Hawking radiation). The first task, then, is to calculate the
late-time history of the mirror, which coincides with the worldine of the
star's center written in Eddington-Finkelstein null coordinates.

The calculation is reasonably straightforward if one carries out a few
intermediate steps in Kruskal coordinates.  Unfortunately, Kruskal
coordinates become ill-defined on the horizon in the extremal limit.
Nevertheless, we have found a generalization of the Kruskal transformation
that allows us to calculate the late-time worldline of the star's centre
all the way to the horizon. Remarkably, this trajectory turns out to
correspond to that of a uniformly accelerated mirror in Minkowski
spacetime. The particle spectrum for the uniformly accelerating mirror can
be found by calculating the Bogoliubov coefficients, which relate the In
and Out vacua. These are known to be proportional to the $K_1$ Bessel
function \cite{FD77,Birrell82}. Consequently, the spectrum is
non-Planckian, and no temperature can be assigned to the object at any
time during its history, in contrast to the situation regarding
non-extremal incipient black holes.

Apart from its nonthermal character, the most striking aspect of the
extremal black hole spectrum is that its amplitude contains a lumped
constant $A$ that depends on the history of the collapsing object. At
first sight this appears to be a direct violation of the no-hair theorems.
Furthermore, given that the object is an extremal one ($Q^2 = M^2$),
because it is radiating neutral particles, the mass will decrease relative
to the charge, leaving a naked singularity as the final product of
collapse, in apparent violation of the cosmic censorship hypothesis.

We have, however, examined the behavior of the stress-energy-momentum
tensor and its variance for these extremal incipient black holes. For the
uniformly accelerated mirror, the flux is known to exhibit certain
pathologies: Its expectation value is zero, despite the fact that
particles are definitely being emitted, as demonstrated by the nonzero
Bogoliubov coefficients. That the expectation value of the flux vanishes
saves the cosmic censorship hypothesis because, although particles {\it
are\/} created, the average energy emitted is zero. (This
counter-intuitive result is related to the fact that the $K_1$ Bessel
function, and hence the spectrum, diverges in the zero energy limit, so a
detector is overwhelmed by soft scalar particles.) We find, moreover, that
the variance of the flux, which also contains the constant $A$, is nonzero
but vanishes as a power law at late times. On the one hand, the fact that
the expectation value of the flux is zero and its variance {\it does\/}
vanish at late time saves the no-hair theorem because any measurement of
the object's history becomes progressively more difficult. On the other
hand, that the decay of the variance obeys a power law is a further proof
that extremal black holes are qualitatively different from non-extremal
ones, for which the variance of the flux vanishes exponentially with time.

Although the former results are valid just for incipient black
holes, one can take the as hints towards more general
conclusions, we can then conjecture that {\em all} extremal black
holes are not thermal objects; more precisely, {\em the set of
thermodynamical states of the black holes does not include
extremal ones\/}. An alternative possibility, however, is that
the test field approximation used in all calculations of this
sort is inappropriate when dealing with such delicate issues, and
that one must go to a more sophisticated theory. The divergence
in the particle spectrum of the uniformly accelerated mirror, for
example, is reminiscent of the typical infrared catastrophe of
QED, which manifests itself in the process of
Bremsstrahlung~\cite{mandl}. Just as those divergences are
removed by considering the recoil of electrons, it might be
necessary to consider recoil of the moving mirror under the
action of the scalar field \cite{par95}. Nevertheless, it is far
from obvious that including such radiative corrections for the
mirror could be transplanted in any straightforward way to the
case of an incipient black hole, and it is equally unobvious that
it would transform the Bessel-like spectrum into a thermal one.
Taking backreaction into account would in fact be an admission
that the test-field approximation employed is {\em never\/} valid
and would be tantamount to a confession that a more sophisticated
theory is indeed necessary in order to assess the true meaning of
the third law of black hole dynamics.

It may be that even the inclusion of backreaction will prove insufficient
to impart thermal properties to extremal black holes. Simultaneously with
our work, Anderson, Hiscock and Taylor~\cite{AHT00} have demonstrated that
for static RN geometries, zero-temperature black holes cannot exist if one
considers spacetime perturbations due to the backreaction of quantum
fields. Thus, at the current state of knowledge it appears that either
extremal black holes are simply not thermodynamic objects, or they
represent solutions in which the external field approach to semiclassical
gravity breaks down. In either case they not only differ dramatically from
their non-extremal counterparts but certainly represent the limit of
current theory.

{\it Note added.}  The conclusions discussed here for extremal 
Reissner-Nordstr\" om black holes evidently pertain to extremal Kerr black holes
as well.  See \cite{Rothman00}.

\section*{Appendix: Asymptotic worldlines for incipient
extremal black holes}
 \setcounter{equation}{0}
 \renewcommand{\theequation}{A.\arabic{equation}}

To derive the late-time worldline for the centre of a body
collapsing to an extremal black hole, we start with the
usual RN geometry in the case $Q^2 = M^2$,
\begin{equation}
 {\rm d}s^2 = -\left(1- \frac{M}{r}\right)^2{\rm d}t^2
    + \left(1- \frac{M}{r}\right)^{-2}{\rm
    d}r^2+r^2\, {\rm d}\Omega^2\;,
\label{metric}
\end{equation}
where ${\rm d}\Omega^2$ is the metric on the unit sphere.
This line element, however, is singular on the horizon ($r =
M$), so we want to transform to Kruskal coordinates.
Normally, one does this by first constructing null
coordinates $u = t-r_*$ and $v = t + r_*$, then going to
Kruskal coordinates $\cal U$ and $\cal V$ which are regular
and finite on the horizon. (Here $r_*$ is the usual tortoise
coordinate.) However, the usual Kruskal transformation is
${\cal U}=-e^{-\kappa u}$ and ${\cal V} = e^{\kappa v}$.
Because $\kappa$ vanishes in the limit of extremality, the
Kruskal transformation results in constant values of $\cal
U$ and $\cal V$, which are thus ill-defined in that limit.

This situation can be remedied, however, by a simple
generalization of the Kruskal transformation. If one defines
$\cal U$ and $\cal V$ through
\begin{equation}
\left.
\begin{array}{l}
  u = -\psi(-{\cal U})\\ v = \psi({\cal V})
\end{array}\right\}\;,
\label{Gen}
\end{equation}
where
\begin{equation}
 \psi(\xi)= 4M\left(\ln\xi - \frac{M}{2\xi}\right)\;,
 \label{psi}
\end{equation}
then it is not difficult to show that the extremal RN metric
(\ref{metric}) can be rewritten in terms of ${\cal U}$ and
${\cal V}$ as
\begin{equation}
 {\rm d}s^2 = -\frac{(r-M)^2}{r^2}
 \psi'(-{\cal U})\psi'({\cal V}){\rm d}{\cal U}{\rm d}{\cal V}+r^2\,{\rm d}\Omega^2\;.
\label{Genmetric}
\end{equation}
This metric is apparently degenerate on the horizon, but in
fact near $r = M$ we have ${\cal
U}=-\psi^{-1}\left(-u\right)\sim-\psi^{-1}
\left(\psi(r-M)\right)=-\left(r-M\right)$ and
\begin{equation}
\psi'(-{\cal U})\sim \frac{4M}{r-M}+\frac{2M^2}{(r-M)^2}\sim
\frac{2M^2}{(r-M)^{2}}\;.
\label{psi'}
\end{equation}
Furthermore, since $\cal V$ is everywhere nonzero and finite
then $\psi'({\cal V})$ is regular there. Now the form taken
by the metric (\ref{Genmetric}) is asymptotically
\begin{equation}
{\rm d}s^{2}\sim -\frac{2M^2}{r^{2}}\psi'({\cal V}) {\rm d}{\cal
U}{\rm d}{\cal V}+r^2\,{\rm d}\Omega^2\;,
\label{Finmetric}
\end{equation}
because the $(r-M)^2$ in the numerator of Eq.\ (\ref{Genmetric})
is killed by the $(r-M)^2$ in the denominator of Eq.\
(\ref{psi'}). Consequently, $\cal U$ and $\cal V$ are good
Kruskal-like coordinates.\footnote{Double-null coordinates that
are regular on general degenerate Killing horizons have been
constructed by Lake \cite{lake}.  In the present case, Lake's
coordinates would essentially correspond with keeping only the
last term inside brackets in Eq.\ (\ref{psi}).  We prefer to use
instead the function $\psi$ given by Eq.\ (\ref{psi}) because it
allows us to define coordinates $\cal U$ and $\cal V$ that cover
not only a region near the horizon, but the whole exterior
spacetime.  Thus, our coordinates $\cal U$ and $\cal V$ can be
regarded as a deformation of Lake's double-null coordinates, with
the difference becoming relevant far from the horizon.}

The next step is to use this result to calculate the
asymptotic worldline of the centre of the star. First, we
match the coordinates $\cal U$ and $\cal V$ onto interior
null coordinates, say, $U$ and $V$. In particular, if two
nearby outgoing rays differ by ${\rm d}U$ inside the star,
then they will differ by ${\rm d}U=\beta({\cal U}) {\rm
d}{\cal U}$, with $\beta$ a regular function, outside.
Similarly, ${\rm d}V = \zeta(v) {\rm d}v$, where $\zeta$ is
another regular function. Then if $v=\bar v$ represents the
last ray that passes through the center of the star before
the formation of the horizon, to first order we have ${\rm
d}V = \zeta (\bar v){\rm d}v$, where $\zeta (\bar v)$ is
constant.

We can write near the horizon
\begin{equation}
{\rm d}U = \beta(0)\frac{{\rm d}{\cal U}}{{\rm d}u}{\rm d}u\;.
\end{equation}
Since for the center of the star ${\rm d}U = {\rm d}V =
\zeta(\bar v){\rm d}v$, this immediately integrates to
\begin{equation}
\zeta(\bar v)(v - {\bar v}) = \beta (0){\cal U}(u) =
         -\beta (0) \psi^{-1}\left(-u\right)
        \sim -2 \beta (0)\frac{M^2}{u}\;.
\end{equation}
The last approximation follows from Eq.\ (\ref{psi}) where
$\xi\sim \psi^{-1}(-2M^2/\xi)$ near the horizon.

Thus the late-time worldline for the center of the star is,
finally, represented by the equation
\begin{equation}
v \sim {\bar v} - \frac{A}{u}\;,\quad u\to +\infty\;, \label{traj}
\end{equation}
where $A = 2\beta (0)M^2/\zeta(\bar v)$ is a positive
constant that depends on the dynamics of collapse. This is a
hyperbolic trajectory, equivalent to that of a uniformly
accelerated mirror in two-dimensional Minkowski space. From
this point on, one can use results already in the
literature, which establish that the particle production in
such a scenario is nonthermal.

\end{document}